\documentclass[12pt]{article}

\renewcommand{\thefootnote}{\fnsymbol{footnote}}

\usepackage{setspace}
\usepackage{amsbsy,amssymb,latexsym,amsfonts,amsmath}
\usepackage{mathrsfs}
\usepackage{graphicx}
\usepackage{bm}
\usepackage{here}
\usepackage{comment}
\usepackage{amsmath}
\usepackage{cases}
\usepackage {empheq}
\usepackage{color}
\usepackage{braket}
 \usepackage{multirow}
\usepackage{amsmath}
\numberwithin{equation}{section}
\makeatletter
\def\EqNumText{\refstepcounter{equation}\cdots\tagform@\theequation}%
\makeatother

\newcommand{\bel}[1]{\begin{equation}\label{#1}}                     
\newcommand{\bal}[1]{\begin{eqnarray}\label{#1}}                     
\newcommand{\be}{\begin{equation}}
\newcommand{\ee}{\end{equation}}

\begin{document}
%
%
\begin{titlepage}
\begin{flushright}
\normalsize
~~~~
NITEP 89\\
OCU-PHYS 531\\
Jan 26, 2021 \\
\end{flushright}

\vspace{15pt}

\begin{center}
{\LARGE Marginal deformations of heterotic interpolating models and exponential suppression of the cosmological constant} \\
\end{center}

\vspace{23pt}

\begin{center}
{ H. Itoyama$^{a, b,c}$\footnote{e-mail: itoyama@sci.osaka-cu.ac.jp},
  Sota Nakajima$^b$\footnote{e-mail: sotanaka@sci.osaka-cu.ac.jp}   }\\

%
\vspace{10pt}
%

$^a$\it Nambu Yoichiro Institute of Theoretical and Experimental Physics (NITEP),\\
Osaka City University\\
\vspace{5pt}

$^b$\it Department of Mathematics and Physics, Graduate School of Science,\\
Osaka City University\\
\vspace{5pt}

$^c$\it Osaka City University Advanced Mathematical Institute (OCAMI)

\vspace{5pt}

3-3-138, Sugimoto, Sumiyoshi-ku, Osaka, 558-8585, Japan \\

\end{center}
%
\vspace{15pt}
\begin{center}
Abstract\\
\end{center}
Following our previous work of 1905.10745 [hep-th], 2003.11217 [hep-th], we study heterotic interpolating models
$D$ dimensionally compactified with constant background fields that include the full set of Wilson lines and radii.
Focusing on the phenomenoloically viable supersymmetry restoring parameter region, we analyze the pattern of gauge symmetry enhancement and the representation of massless fermions.
We obtain the set of cases with the exponentially small cosmological constant.
Our analysis does not depend on non-supersymmetric endpoint models of interpolations.
A part of the moduli space of interpolating models is in one-to-one correspondence with the counterpart of toroidal compactification of heterotic superstrings.

 

\vfill

\end{titlepage}

\renewcommand{\thefootnote}{\arabic{footnote}}
\setcounter{footnote}{0}


\section{Introduction}
Quest of unification of electroweak and strong forces and the experimental discovery of broken supersymmetry have been major driving forces of  some of our high energy theoretical and  experimental activities for more than thirty years.
Currently, the presence of dark energy in our universe obtained from the latest observation
presents a provocative question of how to generate a small and non-vanishing cosmological constant from physics at string scale.

A few years after the introduction of heterotic superstrings\cite{Gross:1984dd},
the modular invariant $SO(16) \times SO(16)$ string theory in ten spacetime dimensions\cite{Dixon:1986iz} was constructed, which is tachyon free and does not possess spacetime supersymmetry.  Following this work, a possibility was suggested in \cite{Itoyama:1986ei} that, upon compactification to lower dimensions (see \cite{Nair:1986zn} for related work), one can construct models nowadays called interpolating models which connect the heterotic superstring and this non-supersymmetric heterotic cousin by the $Z_2$ twist operation as well as by a radius parameter and that, in the supersymmetry restoring region,
one can suppress the value of the cosmological constant. This compactification with twist generalizes the role played by the $T$ duality, which is originally self-dual for the bosonic string\cite{Kikkawa:1984cp}, and the original picture of the heterotic vacua upon toroidal compactification\cite{New Heterotic String Theories in Uncompactified Dimensions $<$ 10}.

Twisted compactifications\cite{Spontaneous Breaking of Supersymmetry Through Dimensional Reduction} of non-supersymmetric heterotic string theory
have now been developed in a variety of directions including $S$ duality\cite{Blum:1997gw,Abel:2018zyt} and more phenomenological studies\cite{Itoyama:2019yst,Itoyama:2020ifw,Abel:2015oxa,Faraggi:2009xy,Kounnas:2015yrc} and many of these cases are characterized  as those where  the one-loop cosmological constant
obeys the following formula\cite{Itoyama:1986ei}:
\begin{equation}\label{cc}
\Lambda^{(10-d)} = (n_F-n_B)\xi a^{10-d} + \mathcal{O}(e^{-1/a}),~~~a\approx 0.
\end{equation}
Here $\xi $ is a positive constant computable from the knowledge of the spectrum of  Kaluza-Klein towers alone,
and therefore from  higher-dimensional quantum field theory alone and $n_F$ and $n_B$ represent
the degrees of freedom of massless fermions and massless bosons respectively\footnote{The clean separation of \eqref{cc} into two terms and related field theoretic limits are currently under study by us
	\cite{Itoyama:1987qz}.}.
The modes in which $n_F=n_B$, that is, the cosmological constant is exponentially suppressed at one-loop level are called $super~no\text{-}scale~models$\cite{Kounnas:2015yrc}. Not only interpolating models but also string models with broken supersymmetry in general have been attracting attention in the context of string
phenomenology and swampland program\cite{Hamada:2015ria}.

In two of our recent publication\cite{Itoyama:2019yst,Itoyama:2020ifw},
we  investigated the 9D interpolating models which connect the two of the
heterotic superstrings with $SO(16) \times SO(16)$ heterotic string by the full set of Wilson lines.
The presence of Wilson lines at a generic point of the moduli space breaks the nonabelian gauge group set in 10-dimensional string theory into the product that includes the abelian gauge groups, spoiling the hope for the unification. This is, however, retrieved in string theory itself: gauge symmetry enhancement takes place at the extrema of the one-loop potential on the moduli space. We have managed to determine  a set of points in the moduli space where this in fact occurs  and the nonabelian gauge groups associated with these points.

In this letter, we extend our previous works in one-dimensional compactification to the $D$-dimensional counterpart. We generalize the analysis of gauge symmetry enhancement in interpolating models by focusing on the splitting of Narain lattice. 
In Sect. \ref{section2}, by $D$-dimensional twisted toroidal compactification, we construct heterotic interpolating models that are marginally deformed by the full set of the constant moduli.
In Sect. \ref{section3}, we study the moduli space of the interpolating models in order to identify the massless spectra. We show that the Wilson lines in the interpolating heterotic models can be represented by those in untwisted toroidal heterotic models. Namely, the massless spectra of interpolating models can be specified by using the information of heterotic superstrings with maximal supersymmetry.
Finally, we apply this analysis to two concrete examples of supersymmetric endpoint models and obtain the same results as in our previous works\cite{Itoyama:2019yst,Itoyama:2020ifw}, including the pattern of symmetry enhancement with the exponentially suppressed cosmological constants.

\section{ $D$-dimensional toroidal compactification of general heterotic interpolating models}\label{section2}

In heterotic string models $D$-dimensional toroidal compactified and deformed by the constant moduli, the dimensionless internal momenta are\cite{New Heterotic String Theories in Uncompactified Dimensions $<$ 10}
\begin{equation}
\label{internal momenta}
\begin{split}
	\ell_{L}^{I} &= m^I - w^{i} A_{i}^{I}, \\
	p_{Li}&=  \frac{1}{\sqrt{2}}\left( m\cdot A_{i} +n_{i}+ w^{j}\left( G_{ij}+B_{ij} -\frac{1}{2}A_{i}\cdot A_{j} \right)  \right) ,\\
	p_{Ri}&=  \frac{1}{\sqrt{2}}\left( m\cdot A_{i} +n_{i}- w^{j}\left( G_{ij}-B_{ij}  +\frac{1}{2}A_{i}\cdot A_{j} \right)  \right) ,
	\end{split}
\end{equation}
where $i=10-D, \cdots, 9$ and $I=1, \cdots, 16$. We shall consider the heterotic models with $D$ compactified dimensions, in which only $X^9$-direction is twisted. We should here introduce a momentum lattice defined as
\begin{align}
	\Lambda\left[\Gamma;\alpha,\beta \right] 
	\equiv
	\left( \eta \bar{\eta}\right) ^{-D}\eta^{-16}\sum_{m^I\in \Gamma}\sum_{w^9\in\boldsymbol{Z}+\alpha}\sum_{n_9\in 2\left( \boldsymbol{Z}+\beta\right) }\sum_{w^{i\neq 9}, n_{i\neq 9}\in\boldsymbol{Z}}  q^{\frac{1}{2}\left( |\ell_{L}|^2 + p_{L}^2 \right) } \bar{q}^{\frac{1}{2} p_{R}^2},
\end{align}
where  $\Gamma$ is a 16-dimensional Euclidean lattice.

In order to construct an interpolating model, we start from a heterotic string model compactified on $T^D$ with maximal supersymmetry:
\begin{align}\label{Z++}
Z^{+}_{+}=Z_{B}^{(8-D)}\left( \bar{V_8}-\bar{S_8}\right) \sum_{\beta=0,1/2}
\Lambda\left[\Gamma_{16};0,\beta\right] ,
\end{align}
where $\Gamma_{16}$ is a 16-dimensional Euclidean even self-dual lattice and $Z_{B}^{(8-D)}=\tau_{2}^{\frac{8-D}{2}}\left( \eta \bar{\eta} \right)^{-(8-D)} $ is the contribution from the bosonic part propagating in $(10-D)$-dimensional spacetime.

 Let $\mathcal{T}^{(9)}$ and $Q$ be operators representing half translations around $X^{9}$-direction and $X^{I}$-directions respectively. By using a shift vector $\delta^{I}\in\frac{1}{2}\Gamma_{16}$, the operator $Q$ can be represented by $\exp\left(  2\pi i m\cdot \delta\right)$ for $m^{I}\in\Gamma_{16}$, and we can split $\Gamma_{16}$ into $\Gamma_{16}^{+}$ and $\Gamma_{16}^{-}$ which are respectively even and odd under $Q$:
\begin{align}\label{Gammapm}
\Gamma_{16}^{+}=\left\lbrace \left.  m^{I}\in \Gamma_{16} ~\right| \delta \cdot m \in \boldsymbol{Z} \right\rbrace,~~~~~
\Gamma_{16}^{-}=\left\lbrace \left.  m^{I}\in \Gamma_{16} ~\right| \delta \cdot m \in \boldsymbol{Z} +\frac{1}{2}\right\rbrace .
\end{align}
An interpolating model is constructed by orbifolding $Z^{+}_{+}$ by the $\boldsymbol{Z}_2$ action $\alpha=(-1)^{F}Q\mathcal{T}^{(9)}$, where $F$ is the spacetime fermion number. The action of $\alpha$ on $Z^{+}_{+}$ gives
\begin{align}\label{Z+-}
Z^{+}_{+}\xrightarrow{\alpha}Z^{+}_{-}=Z_{B}^{(8-D)}\left( \bar{V_8}+\bar{S_8}\right) \sum_{\beta=0,1/2} e^{2\pi i\beta}
\left( \Lambda\left[\Gamma^{+}_{16};0,\beta\right] - \Lambda\left[\Gamma^{-}_{16};0,\beta\right] \right) .
\end{align}
For the partition function to be modular invariant, we must add the twisted sectors:
\begin{align}\label{Z-+}
&Z^{+}_{-}\left( -1/\tau\right)=Z^{-}_{+}\left( \tau\right) =Z_{B}^{(8-D)}\left( \bar{O_8}-\bar{C_8}\right) \sum_{\beta=0,1/2}
\Lambda\left[\Gamma_{16}+\delta;0,\beta\right].\\
&Z^{-}_{+}\xrightarrow{\alpha}Z^{-}_{-}=Z_{B}^{(8-D)}\left( \bar{O_8}+\bar{C_8}\right)\sum_{\beta=0,1/2} e^{2\pi i\beta}
\left( \Lambda\left[\Gamma^{+}_{16}+\delta;0,\beta\right] - \Lambda\left[\Gamma^{-}_{16}+\delta;0,\beta\right] \right),
\end{align}
where we assume that $\delta^{2}$ is an integer.
As a result, the total partition function is 
\begin{align}\label{Ztotal}
Z^{(10-D)}_{int}&=\frac{1}{2}\left( Z^{+}_{+}+Z^{+}_{-}+Z^{-}_{+}+Z^{-}_{-}\right) \nonumber\\
&=Z_{B}^{(8-D)}\left\lbrace \bar{V_8}\left( \Lambda\left[\Gamma_{16}^{+};0,0 \right]+\Lambda\left[\Gamma_{16}^{-};0,1/2 \right] \right)\right. \nonumber \\
&~~~~~~~~~~~~
\left. -\bar{S_8}\left( \Lambda\left[\Gamma_{16}^{+};0,1/2 \right]+\Lambda\left[\Gamma_{16}^{-};0,0 \right] \right)\right. \nonumber\\
&~~~~~~~~~~~~
\left. +\bar{O_8}\left( \Lambda\left[\Gamma_{16}^{+}+\delta;1/2,0 \right]+\Lambda\left[\Gamma_{16}^{-}+\delta;1/2,1/2 \right] \right)\right. \nonumber\\
&~~~~~~~~~~~~
\left. -\bar{C_8}\left( \Lambda\left[\Gamma_{16}^{+}+\delta;1/2,1/2 \right]+\Lambda\left[\Gamma_{16}^{-}+\delta;1/2,0 \right] \right)
\right\rbrace .
\end{align}
In the limit where the volume of the compact space goes to zero, we can check that $Z_{int}^{(10-D)}$ provides a 10D non-supersymmetric endpoint model whose partition function is
 \begin{align}
 Z^{(10)}_{M_2}=Z_{B}^{(8)}\left\lbrace \bar{V_8} \Lambda^{(16)}\left[ \Gamma_{16}^{+} \right] -\bar{S_8} \Lambda^{(16)}\left[ \Gamma_{16}^{-}\right]+\bar{O_8} \Lambda^{(16)}\left[ \Gamma_{16}^{+}+\delta\right] -\bar{C_8} \Lambda^{(16)}\left[ \Gamma_{16}^{-}+\delta\right] 
 \right\rbrace,
 \end{align}
 where  $ \Lambda^{(16)}\left[ \Gamma \right]=\eta^{-16}\sum_{m^I\in \Gamma} q^{\frac{1}{2}|m|^2 }$.
 As the partition function \eqref{Ztotal} reproduces that of an original 10D supersymmetric model in the large volume limit, we can check that the model constructed above interpolates from a 10D supersymmetric endpoint model to a 10D non-supersymmetric one. There are various interpolations, which depend on the choice of the shift vector $\delta^{I}$\cite{Dixon:1986iz,Nair:1986zn}.

In this letter, we are interested in a part of the moduli space where supersymmetry is asymptotically restoring because the cosmological constant can be exponentially suppressed in this region. The states with $w^{9}\neq 0$ acquire huge mass in this region, so the twisted sectors, in which $\alpha=1/2$, are suppressed. We shall henceforth focus only on the contributions from the untwisted sectors.

\section{Pattern of symmetry enhancement}\label{section3}

In this section, we study the massless spectra of interpolating models constructed in the previous section. The spectra of string theories with some compactified dimensions can be classified into two sectors; sector 1 does not depend on the moduli and sector 2 does. In sector 1, the massless states are at the origin of the momentum lattice, i.e. $\ell_{L}^{I}=p_{L}^{i}=p_{R}^{i}=0$, and their degree of freedom is $8\times \left( 8+16 \right) $, which corresponds to a 10D gravity multiplet and 10D gauge bosons of the Cartan subalgebra of a gauge symmetry whose rank is 16. Since the origin of $\Gamma_{16}$ is even under $Q$, there is no massless fermion in sector 1 in interpolating models. Thus, the special points in the moduli space, where some states in sector 2 are massless, can realize $n_{F}=n_{B}$. In sector 2, the massless states must satisfy the following two conditions:
\begin{align}\label{masslass condition1}
|\ell_{L}|^{2}+p_{L}^{2}=2,~~~~p_{R}^{2}=0.
\end{align}
Using eq. \eqref{internal momenta} and the invariant Lorentzian inner product $|\ell_{L}|^{2}+p_{L}^{2}-p_{R}^{2}=|m|^2+2n_{i}w^{i}$, these conditions are expressed as
\begin{align}\label{massless conditions}
|m|^2+2n_{i}w^{i}=2,~~~~m\cdot A_{i} - w^{j}\left( G_{ij}-B_{ij} +\frac{1}{2}A_{i}\cdot A_{j} \right) =-n_{i}.
\end{align}
At generic points in the moduli space, sector 2 contains no massless states and the gauge symmetry is $U(1)^{16+2D}$.
However, there are special points in the moduli space where additional massless states appear if conditions \eqref{massless conditions} are satisfied.

Before we study the pattern of symmetry enhancement, it is worth to comment on the shift symmetry in the moduli space.  Let us consider a shift of a Wilson line $A_{k}^{I}$ by a vector $a_{k}^{I}$ accompanied with a shift of $B_{ki}$ by $\frac{1}{2}a_{k}\cdot A_{i}$:
\begin{align}\label{shift}
A_{k}^{I}\to A_{k}^{I}+ a_{k}^{I}, ~~~~B_{ki}\to B_{ki} + \frac{1}{2} a_{k}\cdot A_{i}~~\left(B_{ik}\to B_{ik} - \frac{1}{2} a_{k}\cdot A_{i} \right),
\end{align}
If $k\neq 9$ and $a_{k}^{I}\in \Gamma_{16}^{+}$, then we can check from eq. \eqref{internal momenta} that the partition function \eqref{Ztotal} is invariant with the replacements
\begin{align}\label{redefinition}
m^{I}\to m^{I}-w^{k}a_{k}^{I},~~~~n_{k}\to n_{k}-\frac{1}{2}w^{k}|a_{k}|^{2}+m\cdot a_{k},~~~~n_{i\neq k}\to n_{i\neq k},~~~~ w^{i}\to w^{i}.
\end{align}
In fact, the shift \eqref{shift} is a part of the duality symmetry of heterotic strings on $T^D$\cite{Giveon:1988tt}\footnote{In non-supersymmetric heterotic models, the shift vector $a_{k}^{I}$ has to be an element of $\Gamma_{16}^{+}$, not $\Gamma_{16}$, in order to maintain the splitting $\Gamma_{16}=\Gamma_{16}^{+}+\Gamma_{16}^{-}$.}. On the other hand, for $k=9$, which is the direction with a twist, the partition function is invariant if $a_{k}^{I}\in 2\Gamma_{16}$ rather than $a_{k}^{I}\in \Gamma_{16}^{+}$, as $w_{9}\in \boldsymbol{Z}+\alpha$ and $n_{9}\in 2\left(\boldsymbol{Z}+\beta \right)$. By the definition \eqref{Gammapm} of $\Gamma^{+}_{16}$, it is clear that $v\in \Gamma^{+}_{16}$ if $v\in 2\Gamma_{16}$.

In this letter, for simplicity, we focus on the states in sector 2 with $w^{i}=0$ for all directions not only for $i=9$. Namely, the massless spectra depend only on the value of the Wilson lines $A_{i}^{I}$, and $G_{ij}$ and $B_{ij}$ take generic values. Then, the conditions \eqref{massless conditions} are 
 \begin{align}\label{massless conditions2}
 |m|^2=2,~~~~m\cdot A_{i}  =-n_{i}.
 \end{align}
Note that if an even self-dual lattice $\Gamma_{16}$ is the root lattice of $Spin(32)/\boldsymbol{Z}_2$ (i.e. $\Gamma_{16}=\Gamma^{(16)}_{g}+\Gamma^{(16)}_{s}$), the elements with $|m|^2=2$ are only in $\Gamma_{g}^{(16)}$. We briefly review the conjugacy classes and the characters of $SO(2n)$ in Appendix \ref{appendixA}.

First, we consider conditions \eqref{massless conditions2} in maximally supersymmetric models compactified on untwisted tori. Then, we investigate the moduli space of interpolating models compactified on twisted tori. We shall provide the relation between the Wilson lines in maximally supersymmetric models and those in interpolating models.
\begin{enumerate}
\item Maximally supersymmetric heterotic models\label{case1}\\
The partition function takes the form \eqref{Z++}, and $m^{I}\in \Gamma_{16}$ and $n_{i}\in \boldsymbol{Z}$ for both spacetime vectors and spinors. 
 Denoting as $A^{(g')}_{i}$ the Wilson lines realizing a semisimple subalgebra $g'\subset g$, where $g$ is $SO(32)$ or $E_{8}\times E_{8}$, as a nonabelian part of a gauge symmetry, $A^{(g')}_{i}$ should satisfy\footnote{The rank of $g'$ is not always 16. If it is smaller than 16, there is the product of $U(1)$ in the gauge symmetry, in addition to the semisimple algebra $g'$.}
\begin{equation}\label{AiGtorus}
\begin{split}
\begin{cases}
m\cdot A_{i}^{(g')} \in\boldsymbol{Z}&~~\text{for}~m^{I}\in \Gamma_{g'}~\text{and}~|m|^2=2\\
m\cdot A_{i}^{(g')} \notin\boldsymbol{Z}&~~\text{for}~m^{I}\in  \Gamma_{g}\backslash \Gamma_{g'}~\text{and}~|m|^2=2,
\end{cases}
\end{split}
\end{equation}
where $\Gamma_{g'}$ and $\Gamma_{g}$ are the root lattices of $g'$ and $g$. For example, the condition that the gauge symmetry is maximally enhanced, i.e. $g'=g$, is 
$A_{i}^{(g)I}\in \Gamma^{*}_{g}$, 
where $\Gamma^{*}_{g}$ is the weight lattice of $g$.
From the replacements \eqref{redefinition}, we find that the interesting sector, in which $w^{i}=0$ and $m^{I}\in \Gamma_{g}$, is invariant under the shift \eqref{shift} with $a_{k}^{I}\in \Gamma^{*}_{g}$, so the condition $A_{i}^{(g)I}\in \Gamma^{*}_{g}$ means $A_{i}^{I}=\left( 0^{16}\right) $ up to the shift symmetry\footnote{Including the $w^{i}\neq 0$ sector, the models are invariant under the shift \eqref{shift} with $a_{k}^{I}\in \Gamma_{16}$. If $a_{k}^{I}\in \Gamma_{v}$ or $a_{k}^{I}\in \Gamma_{c}$, the Wilson lines can not vanish by the shift symmetry and there are possibilities to give the gauge symmetries larger than $SO(32)$, e.g. $SO(34)$.}\label{footnote}.

\item Interpolating heterotic models\\
\begin{table}[t]
	\centering
	\begin{tabular}{|c||c|c|c|} \hline
		& $n_{i\neq 9}$ & $n_9$& $m^I$ \\ \hline
		\multirow{2}{*}{vector} &$\boldsymbol{Z}$  &$2\boldsymbol{Z}$ & $\Gamma_{16}^{+}$\\\cline{2-4}
		&$\boldsymbol{Z}$  & $2\boldsymbol{Z}+1$&$\Gamma_{16}^{-}$ \\\hline
		\multirow{2}{*}{spinor} & $\boldsymbol{Z}$ & $2\boldsymbol{Z}$&$\Gamma_{16}^{-}$ \\\cline{2-4}
		& $\boldsymbol{Z}$ & $2\boldsymbol{Z}+1$&$\Gamma_{16}^{+}$ \\ \hline
	\end{tabular}
\caption{Assignment of $n_{i}$ and $m^{I}$ in interpolating models}
	\label{table}
\end{table}
In order to distinguish from the maximally supersymmetric case, we express the Wilson lines in interpolating models as $\tilde{A}_{i}^{I}$.
From the partition function \eqref{Ztotal}, vectors and spinors have the values of $n_{i}$ and $m^{I}$ following Table \ref{table}.
Conditions \eqref{massless conditions2} for the $X^{i\neq 9}$-directions are the same as in the maximally supersymmetric case since $n_{i\neq 9}\in \boldsymbol{Z}$ for any $m^{I}$.
On the other hand, for the $X^{9}$-direction, the value of $n_{9}$ depends on the splitting of $\Gamma_{16}$. So, the second condition in \eqref{massless conditions2} for massless vectors is
\begin{align}\label{A9,masslessvector}
\tilde{A}_{9}\cdot m \in 2\boldsymbol{Z}~~\text{for}~ m^{I}\in\Gamma_{16}^{+}~~\text{and/or}~~\tilde{A}_{9}\cdot m \in 2\boldsymbol{Z}+1~~\text{for}~m^{I}\in\Gamma_{16}^{-},
\end{align}
while for massless spinors,
\begin{align}\label{A9,masslessspinor}
\tilde{A}_{9}\cdot m \in 2\boldsymbol{Z}+1~~\text{for}~ m^{I}\in\Gamma_{16}^{+}~~\text{and/or}~~\tilde{A}_{9}\cdot m \in 2\boldsymbol{Z}~~\text{for}~m^{I}\in\Gamma_{16}^{-}.
\end{align}
From \eqref{A9,masslessvector}, we find that the Wilson lines $\tilde{A}_{i}^{(g')}$ which realize a semisimple Lie algebra $g'$ in interpolating models are represented by using $A_{i}^{(g')}$ as follows:
\begin{align}\label{tildeA}
\tilde{A}_{i\neq 9}^{(g')}=A_{i\neq 9}^{(g')},~~~\tilde{A}_{9}^{(g')}=2\left( A_{9}^{(g')}+\delta\right) .
\end{align}
For instance, the Wilson lines with maximally gauge symmetry enhancement are $\tilde{A}_{i\neq 9}^{I}=\left( 0^{16}\right) $, $\tilde{A}_{9}^{I}=2\delta^{I}$. We can check from the replacemants \eqref{redefinition} that the sector with $w_{i}=0$ and $m^{I}\in \Gamma_{g}$ in interpolating models is invariant under the shift \eqref{shift} of $\tilde{A}_{9}^{I}$ by $a_{9}^{I}\in 2\Gamma_{g}^{*}$, which is consistent with the relations \eqref{tildeA} and the invariance of $A_{9}^{I}$ under the shift \eqref{shift} with $a_{9}^{I}\in \Gamma_{g}^{*}$.

The massless spinors transform in a different representation of $g'$ than the massless vectors if they exist. Conditions \eqref{A9,masslessspinor} and the relations \eqref{tildeA} imply that spinors can be massless if there is a set of $m^{I}\in \Gamma_{g}\backslash \Gamma_{g'}$ satisfying $A_{9}^{(g')}\cdot m\in \boldsymbol{Z}+1/2$. 
\end{enumerate}
The relation \eqref{tildeA} implies that the massless spectra of interpolating models can be identified only from the information of the supersymmetric endpoint models. Namely, the value of the shift vector $\delta^{I}$, which determines the non-supersymmetric endpoint model of the interpolation, is not required in order to clarify the massless spectra of interpolating models.  

\subsection{Example 1: supersymmetric $SO(32)$ endpoint model}

As a example, let us consider the interpolation in which one of the endpoint model is the supersymmetric $SO(32)$ heterotic model. The choice of the model, which is non-supersymmetric, at the other endpoint depends on the shift vector $\delta^I$: for example, the choice $\delta^{I}=\left( \left( \frac{1}{2}\right)^{8}, 0^{8} \right)$
gives the $SO(16)\times SO(16)$ model, which was investigated in \cite{Itoyama:2019yst,Itoyama:2020ifw}.
In the 10D supersymmetric $SO(32)$ model, $\Gamma_{16}$ is the root lattice of $Spin(32)/\boldsymbol{Z}_2$ and the contribution of $X^{I}$ to the partition function is
\begin{align}
\eta^{-16}\sum_{m^I\in \Gamma_{16}} q^{\frac{1}{2}|m|^2 }=O_{32}+S_{32}=O_{16}O_{16}+V_{16}V_{16}+S_{16}S_{16}+C_{16}C_{16}.
\end{align}

As simple configurations, we consider the following Wilson lines $A_{i}$:
\begin{align}\label{wilson lines example1}
A_{i\neq 9}^{I}=\left( 0^{16}\right) ,~~~~A_{9}^{I}=\left( 0^{p}, \left( \frac{1}{2} \right)^{q}, \left( \frac{1}{4} \right)^{r}\right),~~~~~p+q+r=16.
\end{align}
Note that the above configuration is in the maximally supersymmetric model and we can obtain those in the interpolating model by using the relations \eqref{tildeA}.
Obviously $A_{i\neq 9}\cdot m\in \boldsymbol{Z}$ for any $m^{I}\in \Gamma_{16}$, so the massless spectrum of the interpolating model is determined by $A_{9}^{I}$. 
Following the above discussion about the massless spectrum in sector 2, massless vectors and massless spinors have the following values of $m^I$:
\begin{align}\label{massless vectors example}
\text{vectors: }m^I&=\left( \underline{\pm 1 ,\pm 1, 0^{p-2}},0^{q+r}\right),~\left(0^{p}, \underline{\pm 1 ,\pm 1, 0^{q-2}},0^{r}\right),~ \left(0^{p+q}, \underline{+ 1 ,-1, 0^{r-2}}\right),\\
\text{spinors: }m^I&=\left( \underline{\pm 1 , 0^{p-1}}, \underline{\pm 1 , 0^{q-1}},0^r\right),~\pm\left(0^{p+q}, \underline{+ 1 ,+1, 0^{r-2}}\right),
\end{align}
where the underline indicates the permutations of the components. Thus, the gauge symmetry is $SO(2p)\times SO(2q)\times U(r)$\footnote{We have omitted the abelian factors which  come from the vectors $G_{i\mu}$ and $B_{i\mu}$.} and the massless spinors transform in a bi-fundamental representation of $SO(2p)\times SO(2q)$ and an antisymmetric representation and its conjugate of $SU(r)$. At the point \eqref{wilson lines example1} in the moduli space, 
\begin{align}
n_B&=8\left\lbrace  2p(p-1)+2q(q-1)+r(r-1)+24\right\rbrace  ,\\
n_F&=8\left\lbrace  4pq+r(r-1)\right\rbrace ,
\end{align}
where $8\times24$ is the degree of freedom in sector 1.
 With $p+q+r =16$, the solutions of $n_F=n_B$ are $( p, q, r) = (6,6,4),~ (\underline{6,7},3),~(\underline{7,9},0)$.
The comological constant is therefore exponentially suppressed when the gauge symmetry is enhanced to $SO(12)\times SO(12)\times U(4)$ or $SO(14)\times SO(12)\times U(3)$ or $SO(18)\times SO(14)$. 

\subsection{Example 2: supersymmetric $E_8 \times E_8$ endpoint model}

The other example is constructed from the supersymmetric $E_8\times E_8$ heterotic model. In this supersymmetric model, an even self-dual lattice $\Gamma_{16}$ is the root lattice of $E_8 \times E_8$, and the contribution from $X^{I}$ is represented by the $SO(16)$ characters as follows:
\begin{align}
\eta^{-16}\sum_{m^I\in \Gamma_{16}} q^{\frac{1}{2}|m|^2 }=\left( O_{16}+S_{16}\right) \left( O_{16}+S_{16}\right).
\end{align}

One of the simple and non-trivial configurations of the Wilson lines is
\begin{align}\label{wilson lines example2}
A_{i\neq 9}^{I}=\left( 0^{16}\right) ,~~~~A_{9}^{I}=\left( 0^{p}, \left( \frac{1}{2} \right)^{q}; 0^{p'}, \left( \frac{1}{2} \right)^{q'}\right),~~~~~p+q=p'+q'=8.
\end{align}
At the point \eqref{wilson lines example2} in the moduli space, 
\begin{align}
n_B&=8\left\lbrace  2p\left( p-1\right) + 2q\left( q-1\right)+2p'\left( p'-1\right)+2q'\left( q'-1\right)+S_{i}(p,q)+S_{i}(p',q')+24\right\rbrace, \\
n_F&=8\left\lbrace  4pq+4p'q'+S_{h}(p,q)+S_{h}(p',q')\right\rbrace,
\end{align}
where $S_{i}(p,q)$ and $S_{h}(p,q)$ are the degrees of freedom from the spinor conjugacy class of $SO(16)$ and defined as follows:
\begin{align}
S_{i}(p,q)&=\begin{cases}
0&p,q\in 2\boldsymbol{Z}+1\\
64&(p,q)=(2,6),(4,4),(6,2)\\
128&(p,q)=(0,8),(8,0),
\end{cases}\\
S_{h}(p,q)&=\begin{cases}
0&p,q\in 2\boldsymbol{Z}+1~\text{or}~(p,q)=(0,8),(8,0)\\
64&(p,q)=(2,6),(4,4),(6,2).
\end{cases}
\end{align}
The solutions of $n_F=n_B$ are 
$(p,q,p',q')=(\underline{3,5},\underline{4,4,})$,
which realize $SO(6)\times SO(10)\times SO(8)\times SO(8)$ gauge symmetry.

Let us comment on the $w^{i\neq 9}\neq 0$ sector. As mentioned in the footnote \ref{footnote}, including the $w^{i\neq 9}\neq 0$ sector,  we have the possibility of a richer pattern of gauge symmetry enhancement. In particular, the gauge symmetry can be enhanced to a semisimple Lie algebra whose rank is greater than 16 when $G_{ij}$ and/or $B_{ij}$ take special values, which correspond to the fixed points of the duality symmetry. In \cite{Fraiman:2018ebo}, such a pattern of symmetry enhancement has been explored in maximally supersymmetric heterotic strings compactified on $T^D$.

\section*{Acknowledgments}
We thank Shun'ya Mizoguchi and Yuji Sugawara for helpful discussion on this subject.
The work of HI is supported in part by JSPS KAKENHI Grant Number 19K03828
and by the Osaka City University (OCU) Strategic Research Grant 2020 for priority area.


\appendix

\section{Lattices and characters}\label{appendixA}

Irreducible representations of $SO(2n)$ can be classified into four conjugacy classes:
\begin{itemize}
	\item The trivial conjugacy class (the root lattice):
	\begin{align}
	\Gamma^{(n)}_{g}=\left\lbrace \left( n_1,\cdots,n_{n}\right) \left|~ n_{i}\in \boldsymbol{Z},~\sum_{i=1}^{n}n_{i}\in 2\boldsymbol{Z} \right. \right\rbrace .
	\end{align}
	\item The vector conjugacy class:
	\begin{align}
	\Gamma^{(n)}_{v}=\left\lbrace \left( n_1,\cdots,n_{n}\right) \left|~ n_{i}\in \boldsymbol{Z},~\sum_{i=1}^{n}n_{i}\in 2\boldsymbol{Z} +1\right. \right\rbrace .
	\end{align}
	\item The spinor conjugacy class:
	\begin{align}
	\Gamma^{(n)}_{s}=\left\lbrace \left( n_1+\frac{1}{2},\cdots,n_{n}+\frac{1}{2}\right) \left|~ n_{i}\in \boldsymbol{Z},~\sum_{i=1}^{n}n_{i}\in 2\boldsymbol{Z} \right. \right\rbrace .
	\end{align}
	\item The conjugate spinor conjugacy class:
	\begin{align}
	\Gamma^{(n)}_{c}=\left\lbrace \left( n_1+\frac{1}{2},\cdots,n_{n}+\frac{1}{2}\right) \left|~ n_{i}\in \boldsymbol{Z},~\sum_{i=1}^{n}n_{i}\in 2\boldsymbol{Z}+1 \right. \right\rbrace .
	\end{align}
\end{itemize}
The dual lattice of $\Gamma_{g}^{(n)}$ is the weight lattice of $SO(2n)$, which is the set of weights of all conjugacy classes:
\begin{align}
\Gamma^{(n)*}_{g}=\Gamma^{(n)}_{w}=\Gamma^{(n)}_{g}+\Gamma^{(n)}_{v}+\Gamma^{(n)}_{s}+\Gamma^{(n)}_{c}.
\end{align}

Modular invariance of the partition function of the supersymmetric heterotic string theory requires that $X^{I}$ be compactified on an even self-dual Euclidean lattice. In 16-dimensions, only two such lattices exist, one of which is the root lattice of $E_{8}\times E_{8}$:
\begin{align}
\Gamma_{16}=\left( \Gamma^{(8)}_{g}+\Gamma^{(8)}_{s}\right) \times \left( \Gamma^{(8)}_{g}+\Gamma^{(8)}_{s}\right).
\end{align}
The other is the root lattice of $Spin(32)/\boldsymbol{Z}_{2}$ which is the sum of the trivial and spinor conjugacy classes of $SO(32)$:
\begin{align}
\Gamma_{16}= \Gamma^{(16)}_{g}+\Gamma^{(16)}_{s}.
\end{align}

The $SO(2n)$ characters of the corresponding conjugacy classes are defined as
\begin{align}
O_{2n}
&=\frac{1}{\eta^{n}} \sum_{m^{I}\in \Gamma^{(n)}_{g}}q^{\frac{1}{2}|m|^2}
=\frac{1}{2\eta^{n}}\left( \vartheta^{n}
\begin{bmatrix} 
0\\ 
0\\ 
\end{bmatrix}(0,\tau)+ \vartheta^{n}
\begin{bmatrix} 
0\\ 
1/2\\ 
\end{bmatrix}(0,\tau)
\right),\\
V_{2n}
&=\frac{1}{\eta^{n}} \sum_{m^{I}\in \Gamma^{(n)}_{v}}q^{\frac{1}{2}|m|^2}
=\frac{1}{2\eta^{n}}\left( \vartheta^{n}
\begin{bmatrix} 
0\\ 
0\\ 
\end{bmatrix}(0,\tau)- \vartheta^{n}
\begin{bmatrix} 
0\\ 
1/2\\ 
\end{bmatrix}(0,\tau)
\right),\\
S_{2n}
&=\frac{1}{\eta^{n}} \sum_{m^{I}\in \Gamma^{(n)}_{s}}q^{\frac{1}{2}|m|^2}
=\frac{1}{2\eta^{n}}\left( \vartheta^{n}
\begin{bmatrix} 
1/2\\ 
0\\ 
\end{bmatrix}(0,\tau)+ \vartheta^{n}
\begin{bmatrix} 
1/2\\ 
1/2\\ 
\end{bmatrix}(0,\tau)
\right),\\
C_{2n}
&=\frac{1}{\eta^{n}} \sum_{m^{I}\in \Gamma^{(n)}_{c}}q^{\frac{1}{2}|m|^2}
=\frac{1}{2\eta^{n}}\left( \vartheta^{n}
\begin{bmatrix} 
1/2\\ 
0\\ 
\end{bmatrix}(0,\tau)-\vartheta^{n}
\begin{bmatrix} 
1/2\\ 
1/2\\ 
\end{bmatrix}(0,\tau)
\right),
\end{align}
where the Dedekind eta function and the theta function with characteristics are defind as
\begin{align}
\eta(\tau)&=q^{1/24}\prod_{n=1}^{\infty}\left( 1-q^{n}\right),\\
\vartheta
\begin{bmatrix} 
\alpha\\ 
\beta\\ 
\end{bmatrix}(z,\tau)&=\sum_{n=-\infty}^{\infty}\exp\left( \pi i (n+\alpha)^2 \tau +2\pi i (n+\alpha)(z+\beta) \right). 
\end{align}


\begin{thebibliography}{99}
		
		\bibitem{Gross:1984dd} 
		D.~J.~Gross, J.~A.~Harvey, E.~J.~Martinec and R.~Rohm,
		Phys.\ Rev.\ Lett.\  {\bf 54}, 502 (1985).
		
		\bibitem{Dixon:1986iz} 
		L.~J.~Dixon and J.~A.~Harvey,
		Nucl.\ Phys.\ B {\bf 274}, 93 (1986);\\
		L.~Alvarez-Gaume, P.~H.~Ginsparg, G.~W.~Moore and C.~Vafa,
		Phys.\ Lett.\ B {\bf 171}, 155 (1986).
		
		\bibitem{Itoyama:1986ei} 
		H.~Itoyama and T.~R.~Taylor,
		Phys.\ Lett.\ B {\bf 186}, 129 (1987); FERMILAB-CONF-87-129-T, Proceedings of International Europhysics Conference on High-energy Physics, 25 June-1 July 1987. Uppsala, Sweden (C87-06-25).
		
		\bibitem{Nair:1986zn} 
		V.~P.~Nair, A.~D.~Shapere, A.~Strominger and F.~Wilczek,
		Nucl.\ Phys.\ B {\bf 287}, 402 (1987);\\
		P.~H.~Ginsparg and C.~Vafa,
		Nucl.\ Phys.\ B {\bf 289}, 414 (1987).
		
		\bibitem{Kikkawa:1984cp}
		K.~Kikkawa and M.~Yamasaki,
		Phys. Lett. B \textbf{149}, 357-360 (1984)
		;\\
		N.~Sakai and I.~Senda,
		Prog. Theor. Phys. \textbf{75}, 692 (1986)
		[erratum: Prog. Theor. Phys. \textbf{77}, 773 (1987)]
		.
		
		\bibitem{New Heterotic String Theories in Uncompactified Dimensions $<$ 10} K.~S.~Narain,
		Phys.\ Lett.\  {\bf 169B}, 41 (1986);\\
		K.~S.~Narain, M.~H.~Sarmadi and E.~Witten,
		Nucl.\ Phys.\ B {\bf 279}, 369 (1987).
		
		\bibitem{Spontaneous Breaking of Supersymmetry Through Dimensional Reduction} J.~Scherk and J.~H.~Schwarz,
		Phys.\ Lett.\  {\bf 82B}, 60 (1979);
		R.~Rohm,
		Nucl.\ Phys.\ B {\bf 237}, 553 (1984);\\
		C.~Kounnas and B.~Rostand,
		Nucl.\ Phys.\ B {\bf 341}, 641 (1990).
		
			\bibitem{Blum:1997gw} 
		J.~D.~Blum and K.~R.~Dienes,
		Nucl.\ Phys.\ B {\bf 516}, 83 (1998)
		[hep-th/9707160]; Phys.\ Lett.\ B {\bf 414}, 260 (1997)
		[hep-th/9707148].
		
		\bibitem{Abel:2018zyt} 
		S.~Abel, E.~Dudas, D.~Lewis and H.~Partouche,
		JHEP {\bf 1910}, 226 (2019)
		[arXiv:1812.09714 [hep-th]];\\
		H.~Partouche,
		arXiv:1901.02428 [hep-th];\\
		C.~Angelantonj, H.~Partouche and G.~Pradisi,
		arXiv:1912.12062 [hep-th].\\
		S.~Abel, T.~Coudarchet and H.~Partouche,
		arXiv:2003.02545 [hep-th];\\
		T.~Coudarchet and H.~Partouche,
		[arXiv:2011.13725 [hep-th]].
		
		\bibitem{Itoyama:2019yst} 
		H.~Itoyama and S.~Nakajima,
		PTEP {\bf 2019}, no. 12, 123B01 (2019)
		[arXiv:1905.10745 [hep-th]].
		
		\bibitem{Itoyama:2020ifw}
		H.~Itoyama and S.~Nakajima,
		Nucl. Phys. B \textbf{958}, 115111 (2020)
		[arXiv:2003.11217 [hep-th]].
		
		\bibitem{Abel:2015oxa} 
		S.~Abel, K.~R.~Dienes and E.~Mavroudi,
		Phys.\ Rev.\ D {\bf 91}, no. 12, 126014 (2015)
		[arXiv:1502.03087 [hep-th]]; Phys.\ Rev.\ D {\bf 97}, no. 12, 126017 (2018)
		[arXiv:1712.06894 [hep-ph]];\\
		B.~Aaronson, S.~Abel and E.~Mavroudi,
		Phys.\ Rev.\ D {\bf 95}, no. 10, 106001 (2017)
		[arXiv:1612.05742 [hep-th]];\\
		S.~Abel and R.~J.~Stewart,
		Phys.\ Rev.\ D {\bf 96}, no. 10, 106013 (2017)
		[arXiv:1701.06629 [hep-th]].
		
		\bibitem{Faraggi:2009xy}
		A.~E.~Faraggi and M.~Tsulaia,
		Phys.\ Lett.\ B \textbf{683}, 314-320 (2010)
		[arXiv:0911.5125 [hep-th]];\\
		J.~M.~Ashfaque, P.~Athanasopoulos, A.~E.~Faraggi and H.~Sonmez,
		Eur.\ Phys.\ J.\ C {\bf 76}, no. 4, 208 (2016)
		[arXiv:1506.03114 [hep-th]];\\
		A.~E.~Faraggi,
		Eur.\ Phys.\ J.\ C {\bf 79}, no. 8, 703 (2019)
		[arXiv:1906.09448 [hep-th]];\\
		A.~E.~Faraggi, V.~G.~Matyas and B.~Percival,
		arXiv:1912.00061 [hep-th];
		Nucl. Phys. B \textbf{961}, 115231 (2020)
		[arXiv:2006.11340 [hep-th]];
		[arXiv:2010.06637 [hep-th]]; [arXiv:2011.04113 [hep-th]];
		[arXiv:2011.12630 [hep-th]];\\
		A.~E.~Faraggi, B.~Percival, S.~Schewe and D.~Wojtczak,
		[arXiv:2101.03227 [hep-th]].
		
		\bibitem{Kounnas:2015yrc} 
		C.~Kounnas and H.~Partouche,
		PoS PLANCK {\bf 2015}, 070 (2015)
		[arXiv:1511.02709 [hep-th]]; Nucl.\ Phys.\ B {\bf 913}, 593 (2016)
		[arXiv:1607.01767 [hep-th]]; Nucl.\ Phys.\ B {\bf 919}, 41 (2017)
		[arXiv:1701.00545 [hep-th]];\\
		I.~Florakis and J.~Rizos,
		Nucl.\ Phys.\ B {\bf 913}, 495 (2016)
		[arXiv:1608.04582 [hep-th]];\\
		T.~Coudarchet, C.~Fleming and H.~Partouche,
		Nucl.\ Phys.\ B {\bf 930}, 235 (2018)
		[arXiv:1711.09122 [hep-th]];\\
		T.~Coudarchet and H.~Partouche,
		Nucl.\ Phys.\ B {\bf 933}, 134 (2018)
		[arXiv:1804.00466 [hep-th]].\\
		H.~Partouche,
		Universe {\bf 4}, no. 11, 123 (2018)
		[arXiv:1809.03572 [hep-th]].
		
		\bibitem{Hamada:2015ria}
		Y.~Hamada, H.~Kawai and K.~y.~Oda,
		Phys. Rev. D \textbf{92}, 045009 (2015)
		[arXiv:1501.04455 [hep-ph]];\\
		M.~McGuigan,
		[arXiv:1907.01944 [hep-th]];\\
		C.~Angelantonj, Q.~Bonnefoy, C.~Condeescu and E.~Dudas,
		JHEP \textbf{11}, 125 (2020)
		[arXiv:2007.12722 [hep-th]];\\
		B.~S.~Acharya,
		JHEP \textbf{08}, 128 (2020)
		[arXiv:1906.06886 [hep-th]];\\
		B.~S.~Acharya, G.~Aldazabal, E.~Andr\'es, A.~Font, K.~Narain and I.~G.~Zadeh,
		[arXiv:2010.02933 [hep-th]].
		
		\bibitem{Itoyama:1987qz}
		H. Itoyama, Y. Koga and S. Nakajima in progress;\\
		See also H.~Itoyama and P.~Moxhay,
		Nucl. Phys. B \textbf{293}, 685-708 (1987);
		Y.~Arakane, H.~Itoyama, H.~Kunitomo and A.~Tokura,
		Nucl. Phys. B \textbf{486}, 149-163 (1997)
		[arXiv:hep-th/9609151 [hep-th]].
		
		\bibitem{Giveon:1988tt}
		A.~Giveon, E.~Rabinovici and G.~Veneziano,
		Nucl. Phys. B \textbf{322}, 167-184 (1989)
		; Phys. Rept. \textbf{244}, 77-202 (1994)
		[arXiv:hep-th/9401139 [hep-th]].
		
		\bibitem{Fraiman:2018ebo}
		B.~Fraiman, M.~Gra\~na and C.~A.~N\'u\~nez,
		JHEP \textbf{09}, 078 (2018)
		[arXiv:1805.11128 [hep-th]]; \\
		A.~Font, B.~Fraiman, M.~Gra\~na, C.~A.~N\'u\~nez and H.~P.~De Freitas,
		[arXiv:2007.10358 [hep-th]].
		
\end{thebibliography}
\end{document}